\documentclass{iau-JDSS}
\usepackage{graphicx}

\title[short title of paper] 
{Kinematic Properties and Dark Matter Fraction of Virgo Dwarf Early-Type Galaxies}

\author[short author list]   
{E. Toloba$^{1,2}$%
  \thanks{Fulbright Fellow, email: toloba@ucolick.org},
 A. Boselli$^3$, R. Peletier$^4$ \and J. Gorgas$^5$}

\affiliation{$^1$UCO/Lick Observatory, University of California, Santa Cruz, CA 95064\\[\affilskip]
$^2$Observatories of the Carnegie Institution of Washington, Pasadena, CA 91101 \break $^3$Laboratoire d'Astrophysique de Marseille-LAM, 13388 Marseille \break $^4$Kapteyn Astronomical Institute, University of Groningen, the Netherlands \break $^5$Departamento de Astrof\'{i}sica y CC. de la Atm\'{o}sfera, Universidad Complutense de Madrid}

\pubyear{2009}
\volume{Volume 15}  
\pagerange{119--126}
\date{?? and in revised form ??}
\jname{Highlights of Astronomy, Volume 14}
\editors{Ian F Corbett, ed.}
\begin{document}

\maketitle

\begin{abstract}
What happens to dwarf galaxies as they enter the cluster potential well is one of the main unknowns in studies of galaxy evolution. Several evidence suggests that late-type galaxies enter the cluster and are transformed to dwarf early-type galaxies (dEs). We study the Virgo cluster to understand which mechanisms are involved in this transformation. We find that the dEs in the outer parts of Virgo have rotation curves with shapes and amplitudes similar to late-type
galaxies of the same luminosity (Fig. 1). These dEs are rotationally supported, have disky isophotes, and younger ages than those dEs in the center of Virgo, which are pressure supported, often have boxy isophotes and are older (Fig. 1). Ram pressure stripping, thus, explains the properties of the dEs located in the outskirts of Virgo. However, the dEs in the central cluster regions, which have lost their angular momentum, must have suffered a more violent transformation. A combination of ram pressure stripping and harassment is not enough to remove the rotation and the spiral/disky structures of these galaxies. We find that on the the Faber-Jackson and the Fundamental Plane relations dEs deviate from the trends of massive elliptical galaxies towards the position of dark matter dominated systems such as the dwarf spheroidal satellites of the Milky Way and M31. Both, rotationally and pressure supported dEs, however, populate the same region in these diagrams. This indicates that dEs have a non-negligible dark matter fraction within their half light radius.
\keywords{Virgo cluster, dwarf galaxies, kinematics, dynamics, and dark matter.}
\end{abstract}


\begin{figure}[h!]
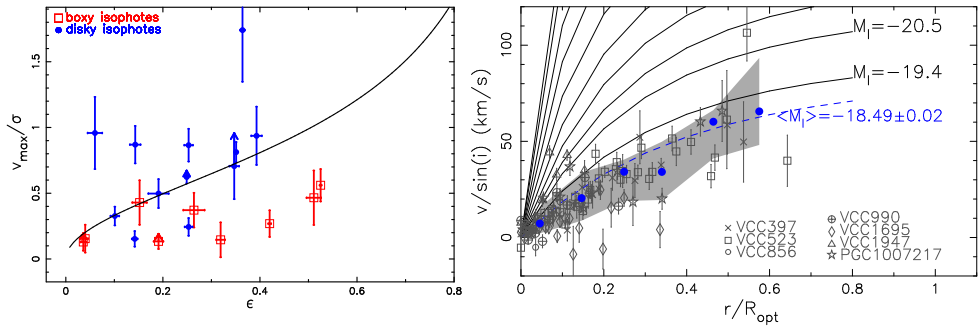

\resizebox{0.45\textwidth}{!}{\includegraphics[angle=-90]{Fig1.ps}}
\resizebox{0.5\textwidth}{!}{\includegraphics[angle=-90]{Fig2.ps}}
  \caption{{\bf Right panel:} Squares indicate dEs with boxy isophotes in the inner ${\rm r} \le 2^{\circ}$ region of the Virgo cluster. Dots show dEs with disky isophotes located in the range 2$^{\circ}\le {\rm r} \le 6^{\circ}$. Rotationally/pressure supported dEs are those above/below the line (model for an isotropic oblate system flattened by rotation), respectively. {\bf Left panel:} Rotationally supported dEs (open symbols, the filled dots represent their median observed rotation curve, 1$\sigma$ deviation in the shaded area) are compared to rotation curves of late-type galaxies (solid and dashed lines).}\label{fig}
\end{figure}





\end{document}